\immediate\write18{makeindex \jobname.nlo -s nomencl.ist -o \jobname.nls} 

 \documentclass[final,3p,times,twocolumn]{elsarticle}

\usepackage{amssymb}
\usepackage[fleqn]{amsmath}
\usepackage{epstopdf}
\usepackage{graphicx}
\usepackage{caption}
\usepackage{subcaption}
\usepackage{float,lscape}
 \usepackage{gensymb}
\usepackage{amsmath}
\usepackage{soul}
\usepackage{color, soul}
\usepackage{siunitx}

    \usepackage{framed} 
    \usepackage{multicol} 

    \usepackage{nomencl} 

    \makenomenclature
    \renewcommand*\nompreamble{\begin{multicols}{2}}
    \renewcommand*\nompostamble{\end{multicols}}
\usepackage{ifthen}

	\renewcommand{\nomgroup}[1]{
           \ifthenelse{\equal{#1}{S}}{\item[\emph{Subscripts}]}{
           \ifthenelse{\equal{#1}{G}}{\item[\emph{Greek letters}]}{                
          \ifthenelse{\equal{#1}{U}}{\item[\emph{Superscripts}]}}}}

\makeatletter
\newcommand*{\rom}[1]{\expandafter\@slowromancap\romannumeral #1@}
\makeatother

\journal{}

\begin{document}

\begin{frontmatter}


\title{Contaminant transport by human passage through an air curtain separating two sections of a corridor: Part I - uniform ambient temperature}

\author[label1,label2]{Narsing~K.~Jha\corref{cor1}}
\ead{navinnaru88@gmail.com}
\author[label1]{\rm D.~Frank}
\author[label2]{P.~F.~Linden}

\address[label1]{~Department~of~Applied~Mathematics ~and~ Theoretical ~Physics,~ University ~of ~Cambridge,~ Wilberforce ~Road,~ CB3~ 0WA ~Cambridge,~ UK}
\address[label2]{~Present~Address:~Department ~of ~ Physics~of ~ Complex~systems,~Weizmann~Institute~of~Science,~ Rehovot,~Israel}

\begin{abstract}

Air curtains are commonly used as separation barriers to reduce exchange flows through an open-door of a building.
Here, we investigated the effectiveness of an air curtain to prevent the transport of contaminants by a person walking
along a corridor from a dirty zone into a clean zone. We conducted small-scale waterbath experiments with fresh
water, brine and sugar solutions, with the brine as a passive tracer for the contaminant in the wake of person. A
cylinder representing human walking was pulled between two fixed points in the channel across the air curtain. We
observed that the air curtain can prevent up to 40\% of the contaminant transport due to the wake of a moving person.
We proposed a new way to evaluate the performance of an air curtain in terms of the deflection modulus and the
effectiveness defined for this iso-density situation, similar to quantities typically used for the case where the fluid
densities in the two zones are different. We observed that the air curtain has an optimal operating condition to achieve
a maximum effectiveness. Dye visualisations and time-resolved particle image velocimetry of the air curtain and the
cylinder wake were used to examine the re-establishment process of the planar jet after its disruption by the cylinder
and we observed that some part of the wake is separated by the re-establishing curtain. We observed that the exchange
flux peaks after the cylinder passes the air curtain and reduces to a typical value after the re-establishment of the
curtain.

\end{abstract}


\end{frontmatter}


\section{Introduction}

Human and vehicular traffic through open doorways is accompanied by a wake which may contain undesirable agents such as airborne contaminants, moisture, odours, insects, heat energy and microorganisms. In the present study, we are interested in the transport of contaminants in the wake of a person walking along an isothermal corridor and ways to minimise it using an air curtain. Such flows can occur in hospital corridors and subway tunnels, and an air curtain can be used to minimise the transmission of airborne diseases in hospitals \citep{adams2011effect,Beggs,Hoffman,Lowbury1971} and to contain smoke in case of fire in a subway tunnel \citep{juraeva2014numerical}. 


In practice, an air curtain is produced by a fan, mounted in a manifold usually located above a doorway, which drives the air and thus establishes a downward planar air jet that acts as a virtual barrier. Air curtains are mainly used in buildings to reduce the buoyancy-driven exchange flows across open doorways between two zones of a building at different temperatures, or, more commonly, between the inside and outside of the building. Under optimal operating conditions, an undisturbed air curtain can reduce the buoyancy-driven air exchange by about 80$\%$ compared to an open doorway \citep{HayStoe1}. 

While turbulent jets are well understood \citep{Gutmark1976, Heskestad1965, rajaratnam1976}, their use as a separation barrier is still being explored. The first fundamental and systematic study on air curtains was carried out by \citet{HayStoe1,HayStoe2}. They identified the deflection modulus $D_m$, defined as the ratio of the jet momentum flux and the pressure difference due to the `stack effect' associated with the buoyancy difference across the doorway,

\begin{equation}
\begin{split}
D_m=\frac{\rho_0 b_0 u_0^2}{g H^2 \left(\rho_d-\rho_l\right)}=\frac{(\rho_0 {Q}_0^2/b_0)}{g H^2 \left(\rho_d-\rho_l\right)}
\\=\frac{{Q}_0^2}{g b_0 H^2 \left(\frac{T_0}{T_d}-\frac{T_0}{T_l}\right)},
\label{eq:Dmdef}
\end{split}
\end{equation}
as the key parameter for the performance of an air curtain. Here, $u_0$, $\rho_0$, $T_0$ and $Q_0$ are the discharge velocity, density, temperature (in Kelvin) and volumetric discharge per unit nozzle length of the jet at the manifold exit, respectively. We approximate the details of this exit as a two-dimensional nozzle of width $b_0$. The height of the doorway is denoted by $H$ and $g$ is the acceleration due to the gravity. For a doorway between two zones at different temperatures the subscripts $d$ and $l$ are used to denote the properties of the dense (cold) and light (warm) air, respectively. The value of $D_m$ determines the stability of an air curtain: a curtain is said to be stable if it reaches the opposite side of the doorway, i.e. the floor if directed downwards, and unstable otherwise \citep{FrankLinden}.

\begin{figure}
    \centering
    \includegraphics[scale=0.5]{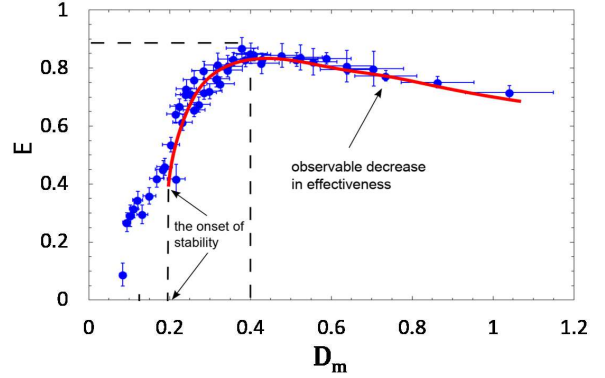}
    \caption{A typical shape of the effectiveness curve $E(D_m)$ as a function of the deflection modulus that was reported in several previous studies. The experimental data have been reproduced from \citep{FrankLinden}.}
    \label{fig:E-sketch}
\end{figure}

For the case $\rho_l\neq\rho_d$, i.e., when there is a horizontal density stratification across the doorway, the performance of an air curtain can be conventionally quantified by the sealing effectiveness $E$, defined as the fraction of the buoyancy-driven exchange flow prevented by the air curtain compared to the open doorway condition,
\begin{equation} \label{eff-1}
E \equiv \frac{q - q_{ac}}{q},
\end{equation}
where $q_{ac}$ and $q$ are the exchange flows through the doorway with and without the air curtain, respectively. A major aim of modelling air curtains is to determine the relationship between $E$ and $D_m$. Typically, as has been reported in several previous studies, the air curtain is unstable for $D_m\lessapprox0.15$ which is reflected in very low and scattered effectiveness values of $E$. For $D_m\gtrapprox0.15$, the air curtain stabilizes and impinges on the floor. $E$ increases with $D_m$ as a result of the air curtain disrupting the organised buoyancy-driven flow through the doorway, until a maximum value of $E$ is reached. The maximum value of $E$ can be as high as 0.8-0.9 and is attained for some value $0.2\lessapprox D_m \lessapprox 0.4$. With further increase in $D_m$, $E$ decreases because of the increased mixing within the air curtain \citep{FrankLinden}. A sketch of a typical $E(D_m)$ curve is presented in figure \ref{fig:E-sketch}.

\citet{choi2012large} and \citet{tang2013different} studied the contaminant transport caused by humans walking through a doorway using large eddy simulation (LES) computations and laboratory experiments, respectively. They examined the effect of a vestibule on the transport and found that a faster walking speed through a vestibule and a larger vestibule results in less mass transport from the contaminated room into the clean room. They also considered the effect of opening various types of doors such as hinged or sliding, and found that the effect of the induced air movement on the human wake had a noticeable effect. \citet{Qi2018} studied the effect of a human presence below an air-curtain (but not human passage) and they observed that the presence of a person either had no influence or reduced the infiltration flux due to blockage of the doorway. 

The aim of the present study is to investigate the effect of an air curtain on the contaminant transport in the wake of a person walking along a corridor in which the air temperature in both sections, and hence density, is uniform. The corridor has a uniform width and height and the air curtain is located at the ceiling and spans the full width of the corridor. Additionally, we also restrict attention to an air curtain blowing air at ambient temperature, so that it is a pure momentum jet and there are no buoyancy-driven flows associated with the air curtain itself. The contaminant transport by human passage through an air curtain in the presence of a buoyancy difference across the air curtain is presented in a companion paper \citep{Jha2020part2}. The study is conducted by means of small-scale waterbath experiments that are dynamically similar to real-scale situations.

The paper is structured as follows. In section \ref{S:2}, we describe the experimental setup and the techniques used for flow visualization, measurements of the contaminant transport and the air curtain effectiveness by means of a conductivity probe, as well as the velocity field measurement using particle image velocimetry (PIV). The experimental results are presented in section \ref{S:3}.
In section \ref{S:3a}, we consider the contaminant transport due to the wake both without and with an operating air curtain and propose new definitions of the deflection modulus $D_{m,c}$ and the air curtain effectiveness $E_c$ in order to evaluate the performance of the air curtain. For flow characterisation, dye visualizations of the jet and the infiltration by the wake are described in section \ref{sec:dye}. Velocity field measurements from PIV showing the interaction of the jet and the wake are presented in section \ref{sec:PIV}, and are used to quantify the temporally varying exchange flux across the air curtain during the entire process of the cylinder motion. Finally, in section~\ref{S:4} we summarise our conclusions.

\section{Experimental methods}
\label{S:2}

\subsection{Setup}

\begin{figure*}
\centering
\includegraphics[scale=0.5]{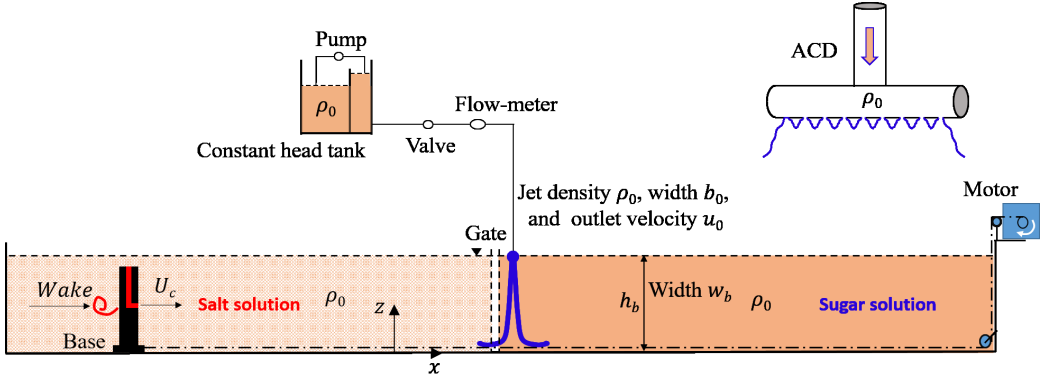}
\caption{Schematic showing the experimental setup. Experiments were conducted in a long channel, one half of which was filled with salt water and the other half with sugar water at the same densities $\rho_d=\rho_l$. Fluid of density $\rho_0=\rho_l=\rho_d$ was supplied to the air curtain device (ACD) from a constant head tank. A vertical gate was installed beside the curtain to separate the fluid when the air curtain device was switched off. The cylinder was driven by a motor, connected by a flexible thread and a system of pulleys. A dye port injecting red dye was attached to the cylinder to track the wake and the infiltration. Blue coloured dye was injected at the nozzle exit to visualise the air curtain.}
\label{fig:ExpSetup}
\end{figure*}

Small-scale waterbath experiments were performed with brine and sugar solutions in a channel representing a corridor with dimensions of length $2L=\SI{2}{\meter}$, width $W=\SI{0.2}{\meter}$, and depth $\SI{0.25}{\meter}$. The channel, a schematic of which is shown in figure \ref{fig:ExpSetup}, was divided in two equal sections by an initially closed vertical gate, which represented a doorway in a corridor. One side of the tank was filled with salt water of density $\rho_d$ and the other side with sugar water of density $\rho_l$. In order to model the situation of a doorway in an isothermal corridor, we adjusted the densities $\rho_l=\rho_d=\SI{1040}{\kilo\gram\per\meter\cubed}$ with the precision of $\SI{0.5}{\kilo\gram\per\meter\cubed}$. Consequently, when the vertical gate was opened there was no buoyancy-driven flux across the doorway between two iso-density sections of the channel. Since molecular diffusion was negligible, sugar and salt acted as passive tracers. All densities were measured using an Anton-Paar density meter DMA 5000 with an accuracy of $\SI{7e-2}{\kilo\gram\per\cubic\meter}$. Additionally, we used a calibrated conductivity probe to measure the salt content $C$ (in $\SI{}{\kilo\gram\per\cubic\meter}$) in each half of the tank.

To represent the passage of a person along the corridor and through the open doorway, a cylinder of diameter $d =\SI{50}{\milli\meter}$ and height $l = \SI{170}{\milli\meter}$ was pulled at a constant velocity $U_c$ along the centreline of the channel from the salt water side to the sugar water side. The base of the cylinder was pulled by a flexible line, attached to a motor via a part of pulleys. To ensure stability during its motion the base of the cylinder consisted of a disc with a larger diameter of $\SI{80}{\milli\meter}$ and height of $\SI{15}{\milli\meter}$. The speed of the cylinder $U_c$ was controlled by the motor and could be varied in the range of $\SI{50} - \SI{250}{\milli\meter\per\second}$. The cylinder was pulled between two fixed points of the channel, namely the midpoints of both channel sections. As we discuss below, the height and diameter of the cylinder were chosen to be representative of a typical adult when compared to the height and width of a typical doorway. The velocity of the cylinder was chosen to ensure that the cylinder motion was dynamically similar to human walking. The length of the experimental tank ensured there was an appropriate distance ($ \sim 10 d$; \citet{Honji1969}) to establish the wake of the cylinder before and after its passage through the doorway.

To study the ability of an air curtain to reduce the contaminant transport due to the cylinder wake, we fitted an air curtain device (ACD) across the top of the tank on the sugar water side next to the vertical gate (figure \ref{fig:ExpSetup}). The ACD consisted of a horizontal cylinder of length $\SI{195}{\milli\meter}$, closed at the ends. The cylinder was filled with fine sponge wrapped in a steel wire mesh to uniformly distribute the flow and was supplied with sugar water of density $\rho_0=\rho_l$ and at a flow rate $q_0$ (in $\SI{}{\milli\meter\cubed\per\second}$) from an overhead tank. A sequence of 39, $\SI{1.0}{\milli\meter}$ diameter holes separated by $\SI{5.0}{\milli\meter}$ (centre to centre) was drilled along the length of the cylinder, facing downwards on the opposite side to the inlet connections. The ACD was positioned in such a way that the holes on the underside of the cylinder were just submerged in water. The circular water jets from the holes merged a few diameters downstream of the nozzle and formed a planar turbulent jet further downstream \citep{Knystautas1964}. The effective planar nozzle width $b_0 = \SI{0.157}{\milli\meter}$ was determined as the ratio of total hole area and length of the cylinder. 

 In a given experiment the volumetric flow rate $q_0$ in $\SI{}{\milli\meter\cubed\per\second}$, measured by the Omega flow meter (FLR1013), was maintained at a constant value with an accuracy of $\pm 3 \%$. We used two values for the volume flux per unit width $Q_0 = q_0/W=\SI{338}{\milli\meter\squared\per\second}$ and $\SI{480}{\milli\meter\squared\per\second}$ (table~\ref{tab1}). The turbulence level at the nozzle exit was not measured but \citet{GuyonnaudSolliec} argued that the turbulence intensity does not noticeably affect the air curtain performance.

To facilitate flow visualisation, we used two dye ports injecting dye of different colours. One dye injection point (blue dye) was placed just underneath the nozzle of the air curtain device on the centreline of the tank. The other dye port (red dye) was attached to the front of the cylinder at half the height of the cylinder. A Nikon D3300 camera was used to capture the  top and the side-view videos of the experiments at 24 f.p.s. {Uniform background illumination was ensured using a projector located  \SI{5}{\meter} behind the channel and then passing it through a translucent film before entering the tank.}

\begin{table*}[!ht]
	\small
	\begin{center}

		\begin{tabular}{lccccccccc}
		\hline
		Experiment & $\rho_d$ & $\rho_l$& $\rho_0$& Air curtain & $Q_0$ &$t_{fw}$& $w_m=l/t_{fw}$ & $U_c$ &$U^*$ \\
		
		& $\SI{}{\kilo\gram\per\meter\cubed}$ & $\SI{}{\kilo\gram\per\meter\cubed}$ & $\SI{}{\kilo\gram\per\meter\cubed}$ && $\SI{}{\milli\meter\squared\per\second}$ &$\SI{}{\second}$ &  $\SI{}{\milli\meter\per\second}$ &  $\SI{}{\milli\meter\per\second}$ & \\
		[3pt]
		\hline
		
		E1 & 1040 & 1040 & 1040 & off & - & - & 0 & 62.5 & 0  \\
		E2 & 1040 & 1040 & 1040 & off & - & - & 0 & 120 & 0  \\
		E3 & 1040 & 1040 & 1040 & off & - & - & 0 & 175 & 0  \\
		E4 & 1040 & 1040 & 1040 & off & - & - & 0 & 192.5 & 0  \\
		E5 & 1040 & 1040 & 1040 & off & - & - & 0 & 210 & 0  \\
		E6 & 1040 & 1040 & 1040 & on  & 338 & 0.56 & 305 & 62.5 &  0.20  \\
		E7 & 1040 & 1040 & 1040 & on  & 338 & 0.56 & 305 & 120 & 0.39  \\
		E8 & 1040 & 1040 & 1040 & on  & 338 & 0.56 & 305 & 175 & 0.57  \\
		E9 & 1040 & 1040 & 1040 & on  & 338 & 0.56 & 305 & 192.5 & 0.63  \\
		E10 & 1040 & 1040 & 1040 & on  & 338 & 0.56 & 305 & 210 & 0.69  \\
		E11 & 1040 & 1040 & 1040 & on  & 480 & 0.39 & 440 & 62.5 & 0.14 \\
		E12 & 1040 & 1040 & 1040 & on  & 480 & 0.39 & 440 & 120 & 0.27 \\
		E13 & 1040 & 1040 & 1040 & on  & 480 & 0.39 & 440 & 175 & 0.40  \\
		E14 & 1040 & 1040 & 1040 & on  & 480 & 0.39 & 440 & 192.5 & 0.44  \\
		E15 & 1040 & 1040 & 1040 & on  & 480 & 0.39 & 440 & 210 & 0.48  \\
		EPIV1 & 998 & 998 & 998 & on & 465 & 0.40 & 420 & 0 & 0 \\
		EPIV2 & 998 & 998 & 998 & on & 465 & 0.40 & 420 & 175 & 0.42 \\
		
		\hline

		\end{tabular}

			
			
		\caption{The experimental conditions and parameter values used in experiments. Here, $t_{fw}$ is the time taken by the curtain to reach the bottom of the tank $z/H=0$ from the cylinder height $z=l$ in a quiescent environment of the same density, $H$ is the height of the curtain from base of the tank, $\rho_l$ and $\rho_d$ are the densities of the fluid in the dense and light side of the tank, respectively.}
		\label{tab1}
	\end{center}
\end{table*}

\subsection{Experimental procedure} 

We first conducted experiments to measure the contaminant transport from the salt water into the sugar water side when the cylinder was moving through an unprotected doorway, i.e., when the air curtain device was switched off. For these experiments (E1 - E5 in table \ref{tab1}), both compartments were initially filled up to the height of $H=\SI{210}{\milli\meter}$. The initial salt content in the salt and the sugar side of the tank was measured using the conductivity probe and labelled as $C_d$ and $C_l$ (which is equal to $0$), respectively. The experiment was started by carefully removing the vertical gate. Subsequently, the cylinder was set into motion from the midpoint of the salt water side at a constant velocity $U_c$ with the direction of travel from the salt water side to the sugar water side and the time measurement started. When the cylinder reached the point at half the distance between the vertical gate and the opposite end wall of the tank (i.e., the midpoint of the sugar water side), the motor driving the cylinder was stopped and the vertical gate was immediately closed. The experiments were conducted for five different cylinder velocities $U_c$ between $\SI{62.5}{\milli\meter\per\second}$ and $\SI{210}{\milli\meter\per\second}$, see table \ref{tab1}. The total time $t$ during which the gate was open and allowed the exchange flow (in the order of magnitude of $\SI{10}{\second}$) was measured using a stop watch with an accuracy of $\SI{0.1}{\second}$. At the end of an experiment, the fluid in each section was fully mixed and the new salt contents in both sides were measured using the conductivity probe, denoted as $C_{dn}$ in the salt water side and $C_{ln}$ in the sugar water side.

The values for the measured salt content were then used to calculate the volume $\tilde V_i$ of salt water infiltrating the sugar water side during an experimental run as

\begin{equation}
     {\tilde V_i = V_{l}\frac{(C_{ln} - C_{l})}{(C_{d} -C_{l})}.}
    \label{eq:Vi-unprotected}
\end{equation}
The infiltration volume $\tilde V_i$ is a measure for the contaminant transport in the cylinder wake. Equation \eqref{eq:Vi-unprotected} is based on the conservation of salt and we implicitly assume that the net flux across the doorway is zero. The initial volume in the sugar water side of the tank is denoted by $V_l$. The error in the measurement of the infiltration volume $\tilde V_i$ is approximately $\pm 2.5\%$ at the highest cylinder speed and $\pm 10\%$ at the lowest speed. We define the time-averaged infiltration flux from the salt water side into the sugar water side due to the wake of the moving cylinder as
\begin{equation}
    {q_{cyl}=\frac{\tilde V_i}{t}.}
    \label{eq:qcyl}
\end{equation}

The experimental procedure for measuring the contaminant transport, i.e., the infiltration volume, when the cylinder was moving through a doorway protected with an operating air curtain was similar but with a few minor differences (E6 - E15 in table \ref{tab1}). As before, the salt water side was filled with brine solution up to the height of $H=\SI{210}{\milli\meter}$ but the water level in the sugar water side was initially set slightly below at $\SI{205}{\milli\meter}$. The experiment was started by switching on the ACD. Once the flow in the jet reached a steady state and the water level in the sugar water side reached the same level $H$ as in the salt water side, the vertical barrier between the two section was opened and the blue dye port switched on. Immediately, the cylinder was set into motion at a constant velocity $U_c$ from the midpoint of the salt water side, and the red dye port was opened to visualise the cylinder wake. We used the same cylinder velocities $U_c$ as for experiments with an unprotected doorway opening. As before, we ended the experiment when the cylinder reached the midpoint of the sugar water side by stopping the motor and closing the vertical gate. We also immediately switched off the flow through the air curtain and closed all the dye ports. After thoroughly mixing the fluid in both sides, we used the conductivity probe to measure the salt contents $C_{ln}$ and $C_{dn}$.

For experiments with an operating ACD, sugar water of density $\rho_0=\rho_l$ was constantly added to the tank at a rate $q_0$, which was the total source volume flux (in $\SI{}{\milli\meter\cubed\per\second}$) of the air curtain. Thus, when calculating the infiltration volume $V_i$ of salt water into the sugar water side a small correction needs to be applied
\begin{equation}
    {V_i = \left(V_l +\beta q_0t\right)\frac{(C_{ln} -C_{l})}{(C_{d} -C_{l})}.}
    \label{eq:Vi-ACD}
\end{equation}
Here, $\beta$ denotes the fraction of the air curtain volume flux that spilled into the sugar water section upon impingement of the air curtain on the channel bottom. For our typical jet nozzle volume fluxes of $q_0=\SI{80000}{\milli\meter\tothe{3}\per\second}$ (corresponding to the flux per unit width of $Q_0=\SI{400}{\milli\meter\tothe{2}\per\second}$) and run-times of less than $\SI{15}{\second}$, the  volume of water added by the curtain to the sugar water compartment was at most $3\%$ of the total volume $V_l$ if $\beta$ = 1. Any value of $\beta$ between 0.5 and 1 can be chosen with an error of less than $3\%$ of $V_{i}$, and we used $\beta$ = 0.5. The time-averaged infiltration flux of salt water into the sugar water side when the ACD is operating is then calculated as
\begin{equation}
    {q_{ac}=\frac{V_i}{t},}
    \label{eq:q_ac}
\end{equation}
where $t$ is the time duration of the door opening in experiment and it is the same for a given cylinder speed $U_c$ both with and without an operating air curtain. We describe the ability of an air curtain to prevent the infiltration flux associated with a moving cylinder in terms of the effectiveness

\begin{equation}
    {E_c=\frac{q_{cyl}-q_{ac}}{q_{cyl}}\left(=\frac{V_i-\tilde V_i}{\tilde V_i}\right).}
    \label{eq:Ec_def}
\end{equation}

We note that although we calculate the infiltration flux $q_{ac}$ as an average quantity for the cylinder motion through the air curtain, in reality, it is expected to be variable with time, especially when the cylinder travels a long distance before and after the air curtain. More precisely, the total infiltration flux $q_{ac}(t)$ can be decomposed as

\begin{equation}\label{eq:QacDecomp}
    {q_{ac}(t)=q_{ac,m}+q_{ac,cyl}(t),}
\end{equation}
where $q_{ac,m}$ is the exchange flux caused by the mixing process within the air curtain, assumed to be time-independent, and $q_{ac,cyl}(t)$ is the infiltration flux due to the curtain-wake interaction. In the absence of a moving cylinder $q_{ac}(t)=q_{ac,m}$. So, were the cylinder started a very long distance away from the air curtain, $q_{ac}(t)$ would have an initial steady value $q_{ac}(t)=q_{ac,m}$. Subsequently, as the cylinder passed underneath the air curtain, the value of $q_{ac}(t)$ would increase by $q_{ac,cyl}(t)$ and then would again return to its initial value $q_{ac,m}$ after the cylinder has travelled an appropriate distance away from the air curtain.

In our experiments, we tried to choose the travelling distance of the cylinder before and after the air curtain appropriately, and focus only on the immediate processes associated with the cylinder passage through the air curtain.  This means that we attempted to limit the travel distance of the cylinder as much as possible and consider the time frame when $q_{ac}(t)$ is increased due to the curtain-wake interaction, i.e., $q_{ac,cyl}(t)\neq 0$. While $q_{ac}(t)$ may still vary during this time frame, it is non-trivial to measure this temporal variation. Thus we measure the time-averaged $q_{ac}$ by measuring the densities at the start and the end of an experiment and use (\ref{eq:Ec_def})  to calculate this `cumulative' effectiveness for a cylinder moving a fixed distance before and after the air curtain. We use PIV measurements described in the next section to discuss whether our choice of the cylinder travelling distances before and after the air curtain is appropriate.

\subsection{{PIV measurements}}

In order to gain insight into the instantaneous flow field and the temporal variation of the infiltration flux, we conducted measurements using the time-resolved 2D particle image velocimetry (PIV) (experiments EPIV1 and EPIV2 in table~\ref{tab1}). Since no tracer for the infiltration flux was required in this set of experiments and in order to avoid the impact of refractive index changes, we used here pure water in both sides of the tank and for the air curtain supply, i.e. $\rho_d=\rho_l=\rho_0=\SI{998}{\kilo\gram\per\meter\cubed}$. The velocity field was measured in the streamwise and vertical plane through the axis of the cylinder. A thin vertical light sheet was generated by passing the light from a pair of 300 W xenon arc lamps through a thin slit. The arc lamp was equipped with paraboloidal dichroic reflectors, which produce about $\SI{35}{\watt}$ of visible light. The light sheet entered the tank after being reflected from a cold mirror to avoid heating the perspex tank. The flow was seeded with $\SI{50}{\micro\meter}$ diameter particles, which were approximately neutrally buoyant. Particle concentrations were chosen such that each correlation box used in the PIV analysis had 6 to 8 particles to provide good correlations. Images were recorded using an ISVI camera with a full resolution of 12 Mpixel. A Nikon $\SI{85}{\milli\meter}$ lens was used and images were captured at a frame rate of 400 f.p.s. with a resolution of 2048$\times$2048 pixels. Exposure times were kept at $\SI{0.8}{\milli\second}$ and no streaking of particles was observed. We performed the analysis of the image pairs in the DigiFlow software using a box size of $33\times33$ pixels. The present PIV measurements are similar to those described in more detail in \citet{olsthoorn2015vortex}.

\subsection{{Dynamical similarity and comparison to real-scale air curtains}}

In the following, we describe the results in terms of the non-dimensional cylinder speed $U^* \equiv U_c/w_m$. We calculate the mean velocity of the air curtain front as $w_m \equiv H/t_{fw}$, where $t_{fw}$ is the time needed by an establishing air curtain to reach the tank bottom $z/H=0$ from the cylinder top height $z=l$, with $z$ being the vertical coordinate. We measured $t_{fw}$ by tracking the air curtain front in the recordings where the blue-dyed air curtain was establishing in the fresh water environment. For the value of the ACD flow rate of $Q_0 \sim \SI{480}{\milli\meter\per\second\squared}$, $w_m$ is about \SI{440}{\milli\meter\per\second}. We varied the cylinder speed from $\SI{62.5}{\milli\meter\per\second}$ to $\SI{210}{\milli\meter\per\second}$, which results in values of $U^*$ from 0.14 to 0.69. 

These values are typical for the human passage through a full-scale air curtain. For example, walking at a moderate pace of $\SI{1}{\meter\per\second} = \SI{3.6}{\kilo\meter\per\hour}$ through a typical air curtain with the speed $w_m \sim \SI{2}{\meter\per\second}$, gives $U^* \sim 0.5$. {The cylinder Reynolds numbers $Re_{cyl} \equiv U_cd/\nu$ are in the range 3,100 -- 10,500 in our experiments compared with $\sim 50,000$ at full scale. Both the experimental and full scale Reynolds numbers are in the shear-layer transition regime \citep{Williamson1996}, so we expect our results to apply at full scale. }

In domestic buildings, doorways are typically between twice as wide as the human waist and up to 5 to 6 times wider in non-domestic buildings, and the door width in our experiments was four times wider than the cylinder diameter $d$. Also, the wake width of the cylinder is around 4$d$ at a distance of $5d$ away from the cylinder axis. Usually, the cylinder crosses a length of approximately $5d$ while the curtain is re-establishing.

\section{Results}
\label{S:3}


\subsection{Infiltration volume and effectiveness of the air curtain}\label{S:3a}

\begin{figure*}[!ht]
\centering\includegraphics[scale=0.45, trim={0.00cm 0 0 0.00cm},clip]{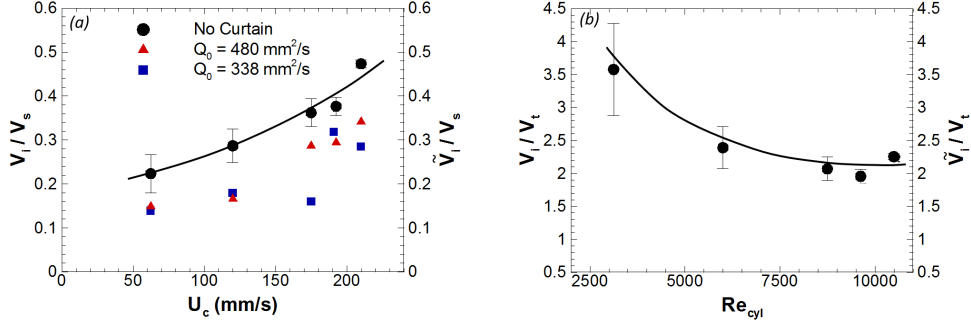}
\caption{ (a) The non-dimensional transported volume $\tilde V_i/V_s$ and $V_i/V_s$ of salt water across the doorway between two iso-density spaces for different cylinder speeds $U_c$ without and with an operating air curtain, respectively. Curves are shown to highlight the trend of the data. (b) Non-dimensional transported volume $\tilde V_i/V_t$ as a function of the cylinder Reynolds number $Re_{cyl}=U_c d/\nu$.} 
\label{fig:V_ent}
\end{figure*}

\begin{figure}[h]
\centering
\includegraphics[scale=0.45]{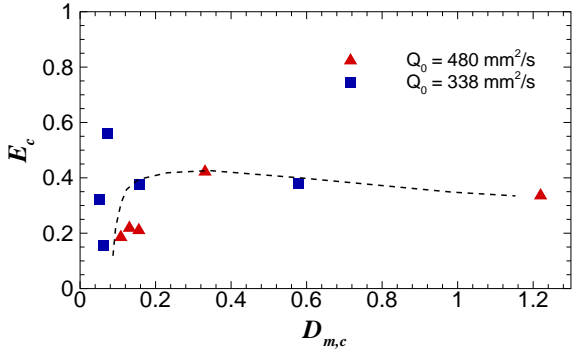}
\caption{The effectiveness curve $E_c(D_{m,c})$ of the air curtain, where the effectiveness $E_c$ and the deflection modulus $D_{m,c}$ in the present case of a cylinder moving between two iso-density rooms are defined by (\ref{eq:Ec_def}) and (\ref{eq:Dmc_def}), respectively. The dashed curve is shown to highlight the trend of the data.}
\label{fig:EcDmc}
\end{figure}

In our experiments, we calculated the transported volume of salt water $\tilde V_i$ and $V_i$ in the wake of the moving cylinder across the doorway using (\ref{eq:Vi-unprotected}) and (\ref{eq:Vi-ACD}), respectively. We non-dimensionalise $\tilde V_i$ and $V_i$ by the swept volume of the cylinder $V_s$ = $U_s l d t = ldL$, where L is the swept length of the cylinder or half length of channel and $t$ is the duration of the experiment. Note that
\begin{equation}
    \frac{\tilde V_i}{V_s}=\frac{\tilde V_i}{U_c ld t}=\frac{q_{cyl}}{U_c ld}
\end{equation}
and
\begin{equation}
    \frac{V_i}{V_s}=\frac{V_i}{U_c ld t}=\frac{q_{ac}}{U_c ld}.
\end{equation}

In figure~\ref{fig:V_ent}a, we plot $\tilde V_i/V_s$ for the case of no air curtain and $\tilde{V}_i/V_s$ for two curtain strengths of $Q_0 = \SI{338}{\milli\meter\squared\per\second}$ and $Q_0 = \SI{480}{\milli\meter\squared\per\second}$ as a function of the cylinder speed $U_c$. We need to plot the data here against $U_c$ rather than $U^*$ since $U^*$ cannot be meaningfully defined when the air curtain is not operating. For a fixed air curtain strength, an increasing $U_c$ corresponds to an increasing $U^*$. When the air curtain is not operating, the infiltration volume $\tilde V_i/V_s$, and hence the contaminant transport, increases with the cylinder speed $U_c$. When the air curtain is switched on, we observe a clear reduction in the transported volume $V_i/V_s$ of salt water across the doorway: the air curtain prevents up to a half of the volume transport. For low cylinder speeds $U_c$, and hence $U^*$ (experiments E6, E7, E11, E12), this reduction is more pronounced and  $V_i/V_s=q_{ac}/(U_c l d)$ rises slightly with the cylinder speed. For higher cylinder speeds $U_c$ (corresponding to higher $U^*$), the infiltration volume $V_i/V_s=q_{ac}/(U_c l d)$ increases noticeably and the data display a noticeable scatter. As we will see in sections \ref{sec:dye} and \ref{sec:PIV}, the infiltration volume $V_i$ is mainly due to the disruption of the air curtain by the moving cylinder and occurs while the air curtain is re-establishing. For higher cylinder speeds relative to a typical air curtain velocity (i.e., for higher values of $U^*$), the air curtain interacts with strong vortices in the cylinder wake during its re-establishment process which leads to the observed scatter in the measured data for $V_i/V_s$. The data for both air curtain strengths are similar. The data for the weaker air curtain ($Q_0 = \SI{338}{\milli\meter\squared\per\second}$, blue squares) are more scattered which reflects the fact that the values of $U^*$ are larger so that the air curtain is more affected by the interaction with the cylinder wake during its re-establishment process. We note here that both air curtain strengths give us values of $U^*$ in a similar range (see table \ref{tab1}). We expect that the measured values of $V_i/V_s$ would differ significantly if we used an air curtain such that, for example, $U^*\gg 1$. We discuss further down below a proper way to non-dimensionalise the data presented in figure \ref{fig:V_ent}a.

We also use the non-dimensionalisation by $V_t = l d U_c\Delta t$, where we take $\Delta t=\SI{1}{\second}$ and $V_t$ is the volume swept by the cylinder in the unit of time. Figure \ref{fig:V_ent}b shows $\tilde V_i/V_t$ as a function of the cylinder Reynolds number $Re_{cyl}=U_c d/\nu$. We observe that $\tilde V_i/V_t$ is independent of the cylinder Reynolds number once the wake is turbulent. This confirms again that our small-scale experiments are dynamically similar to real-scale processes.

As was explained in the introduction, in case of a horizontal density stratification across the doorway, the effectiveness $E$ (see \eqref{eff-1}) of the air curtain is determined as a function of the deflection modulus $D_m$ (see \eqref{eq:Dmdef}). For the case of a cylinder travelling between two iso-density spaces, we introduced the effectiveness of the air curtain in \eqref{eq:Ec_def}.
$E_c$ attains the value of $E_c=1$ if the air curtain completely prevents the infiltration flow due to the cylinder wake and it assumes the value $E_c=0$ if the air curtain has no effect on the flow.

The conventional definition of $D_m$ (see (\ref{eq:Dmdef}) is also not meaningful in case of iso-density rooms, since in that case $D_m\to\infty$. The definition for $D_m$ is the ratio between the initial momentum flux of the air curtain and the horizontal pressure force acting across the doorway due to the stack effect. Motivated by this, we introduce the new deflection modulus $D_{m,c}$ as the ratio of the momentum flux of the air curtain and the lateral momentum flux due to the cylinder motion:

\begin{equation}
    {D_{m,c}=\frac{b_0 u_0^2}{U_c^2 H}\times\frac{H}{l}\times\frac{W}{d}=\frac{W Q_0^2}{b_0 l d U_c^2},}
    \label{eq:Dmc_def}
\end{equation}
where $W$ is the door width (here the full width of the corridor). Large values for $D_{m,c}$ correspond to slow cylinder speeds $U_c$ or to a large air curtain momentum flux (small $U^*$). Small values $D_{m,c}$ describe the situation when the cylinder moves fast and the air curtain momentum flux is small (large $U^*$). Note that the factors $H/l$ and $W/d$ allow the inclusion of the effects of the cylinder geometry in relation to the doorway dimensions. This could be a topic for future investigations but is beyond the scope of the present study.

Figure \ref{fig:EcDmc} shows the measured $E_c(D_{m,c})$ curve for our experiments. From figure \ref{fig:EcDmc} we can deduce the following pattern for the air curtain effectiveness $E_c(D_{m,c})$: we observe low and scattered values of the effectiveness $E_c$ for $D_{m,c}\lessapprox 0.2$. Here, the air curtain is weak compared to the cylinder forcing: it is violently disrupted by the moving cylinder and does not re-establish quickly enough to cut the cylinder wake. As $D_{m,c}$ increases, the air curtain can interrupt the cylinder wake more efficiently and thus $E_c$ increases with $D_{m,c}$. The effectiveness $E_c$ reaches its peak value of $\approx0.4$ for $0.2\lessapprox D_{m,c}\lessapprox 0.4$. For $D_{m,c}\gtrapprox 0.4$, we recognise a slow decrease in the air curtain effectiveness $E_c$. Here, the air curtain is strong compared to the cylinder wake and re-establishes almost immediately after the cylinder passage. However, the self-induced mixing by the air curtain, i.e. $q_{ac,m}$, is large and increasing in this regime so that $E_c$ reduces.

It is instructive to compare the $E_c(D_{m,c})$ curve in figure \ref{fig:EcDmc} with the typical $E(D_m)$ curve in figure \ref{fig:E-sketch} when there is no cylinder motion but the lateral forcing on the air curtain is due to the stack pressure. Both curves display low and scattered effectiveness values for $D_{m,c}\lessapprox0.2$ and $D_m\lessapprox0.2$, respectively. For $D_{m,c}\gtrapprox0.4$ and $D_m\gtrapprox0.4$ we observe a decrease in $E$ and $E_c$, which can be attributed to an increasing $q_{ac,m}$ in both cases. However, the peak value $E_c\approx0.4$ is significantly smaller than the peak value $E\approx 0.8$. The reason for this is that $E(D_m)$ reaches a maximum value when the air curtain completely shields the doorway and the only infiltration flow is due to the entrainment and mixing flux $q_{ac,m}$. For $E_c(D_m)$ curve, in contrast, the air curtain is always temporarily disrupted by the cylinder and $q_{ac,cyl}\neq 0$. We conclude that an optimally operating air curtain is more effective in preventing the stack-pressure driven flux $q$ compared to the equal infiltration flux $q_{cyl}$ associated with the cylinder wake. Similarly, an optimally operating air curtain between two iso-thermal rooms can never reduce the exchange flux $q_{cyl}$ due to the moving cylinder by more than 50\%.

For real-scale air curtain installations, assuming $b_0=\SI{1}{\centi\meter}$, $U_c=\SI{1}{\meter\per\second}$, $l=\SI{1.5}{\meter}$, $W=\SI{2}{\meter}$, $d=\SI{0.5}{\meter}$ and the air curtain discharge velocity in the range $\SI{1}{\meter\per\second}-\SI{5}{\meter\per\second}$ yields values of the modified deflection modulus $D_{m,c} \approx 0.03 -0.66$. This is the range of values that we studied in our small-scale experiments.

\subsection{Dye visualisation}\label{sec:dye}

\begin{figure*}[!ht]
\centering
\includegraphics[scale=0.55]{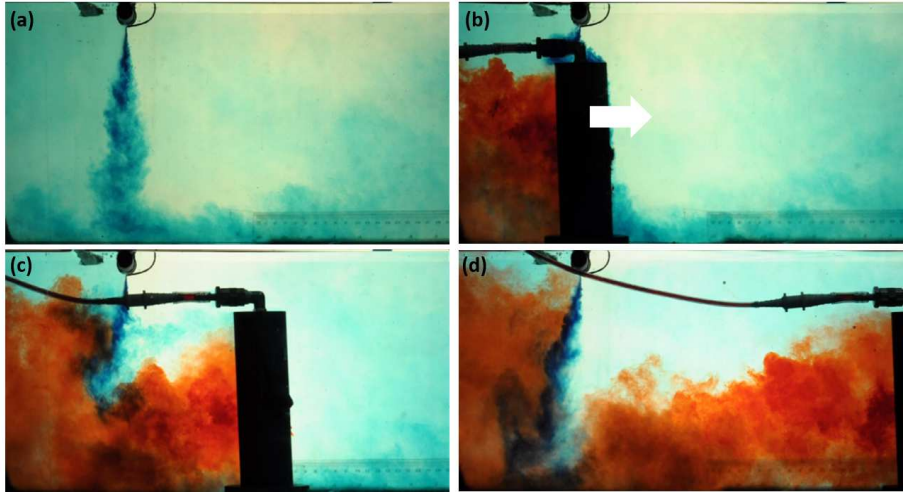}
\caption{Side views of the interaction between the cylinder and the curtain for $U^*= 0.42$, {experiment EPIV2} (see table \ref{tab1}). The cylinder is moving from the salt water to the sugar water side (from left to right as marked in (a)). The blue-dyed curtain jet separates two channel sides in (a) and the red dye visualises the cylinder wake and the infiltration. The infiltration across the curtain and the re-establishment process is shown in (c). The curtain is re-established in (d).}
\label{Fig3}
\end{figure*}


We now present the dye visualisations and use them to further explain the observed flux and effectiveness measurements. Side views of the passage of the cylinder (left to right, marked by the arrow in figure~\ref{Fig3}(b)) are shown in figure \ref{Fig3}. The cylinder was set into motion $\SI{0.5}{\meter}$ away from the air curtain which was not influenced by the cylinder at that distance as can be seen in figure \ref{Fig3}(a). As the cylinder approached the air curtain, there was a small deflection of the air curtain in the direction of motion of the cylinder. This agrees with our definition of $D_{m,c}$ which is based on the momentum flux, or, equivalently, the pressure force that the cylinder exerts on the air curtain. The major disruption occurred as the cylinder passed below the air curtain as shown in figure \ref{Fig3}(b). Immediately after the cylinder passed through the air curtain, the infiltration of salt water (coloured in red) into the sugar water side took place during the period of re-establishment of the curtain. At later times, the air curtain re-established by penetrating the wake as seen in figure \ref{Fig3}(c). Figure~\ref{Fig3}(d) illustrates the moment when the air curtain was first re-established and the infiltrated red fluid can be observed. Quantitative characterisation of the infiltration flux using PIV is presented in the next section.

\subsection{Particle Image Velocimetry}\label{sec:PIV}

\begin{figure}[!t]
\centering\includegraphics[width=0.85\linewidth]{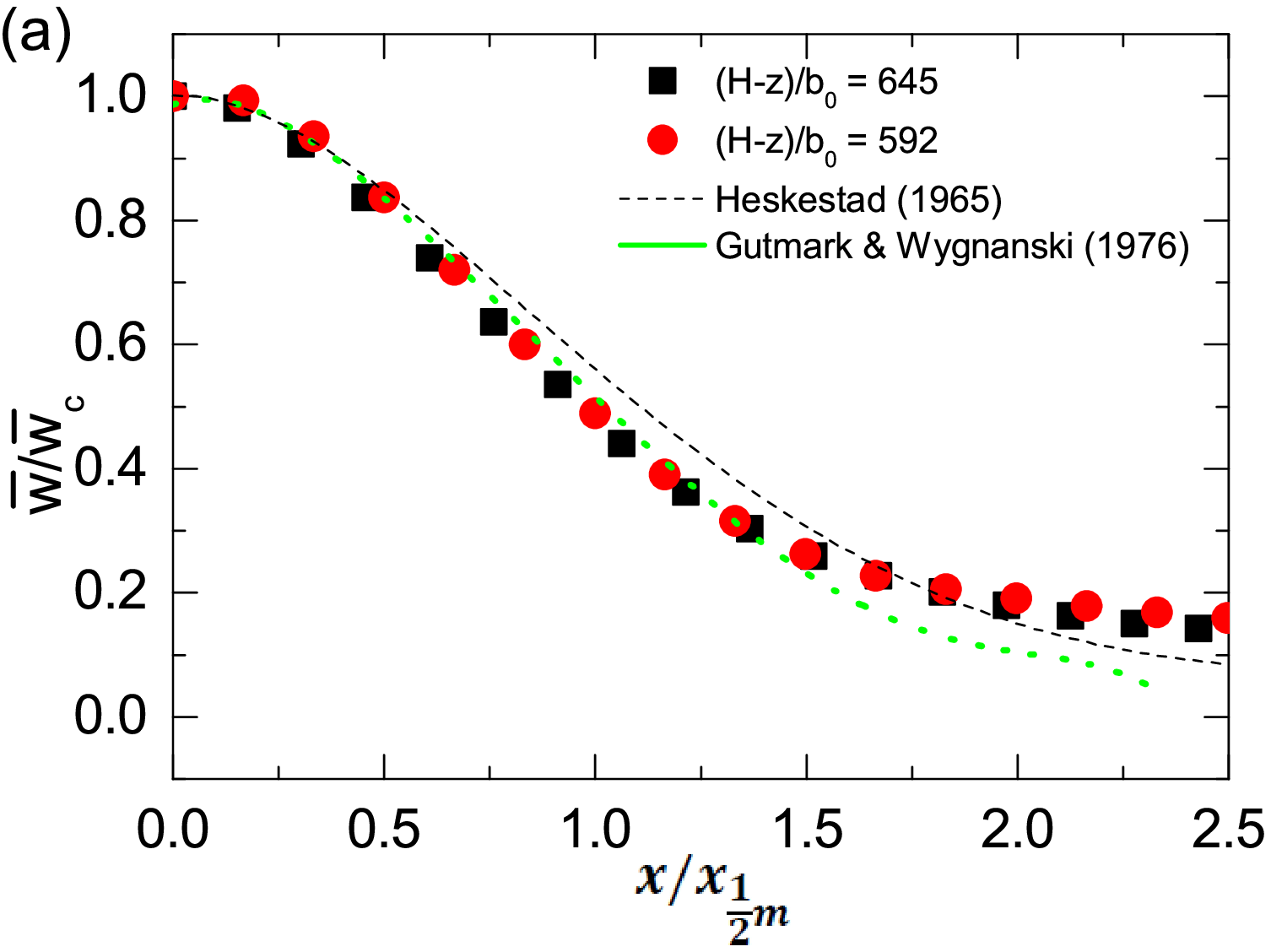}\\
\centering\includegraphics[width=0.85\linewidth]{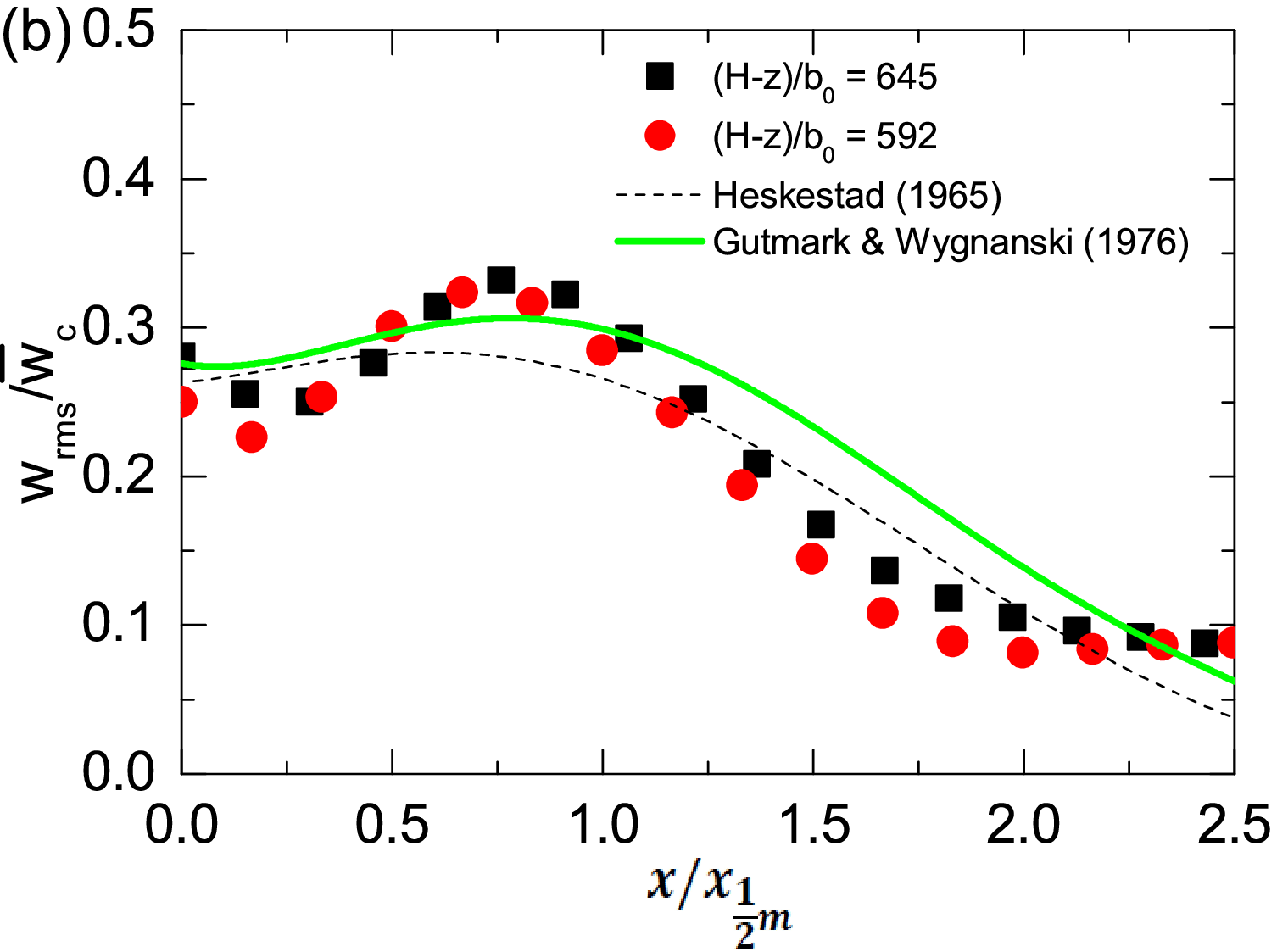}\\
\caption{ Profiles of the mean and fluctuating vertical (axial) velocity of the turbulent jet produced by the air curtain device in the absence of a cylinder wake measured at two downstream locations $(H-z)/b_0 = 592$ and $(H-z)/b_0 =645$ for EPIV1 as described in table~\ref{tab1}.  In (a) the profile of the mean axial velocity, normalised by the centreline velocity $\overline{w}/\overline{w}_c$, is plotted against the normalised lateral distance from the jet axis. In (b), the axial turbulent intensity profile, non-dimensionalised by the mean axial centreline velocity $\overline{w}_c$, is plotted against the normalised lateral distance from the jet axis. For comparison and validation of mean and turbulent properties, the data from \citet{Heskestad1965} and \citet{Gutmark1976} are also shown, with reasonable agreement between their data and ours.}
\label{Fig9aa}
\end{figure}

\begin{figure*}[!ht]
\centering\includegraphics[width=0.85\linewidth]{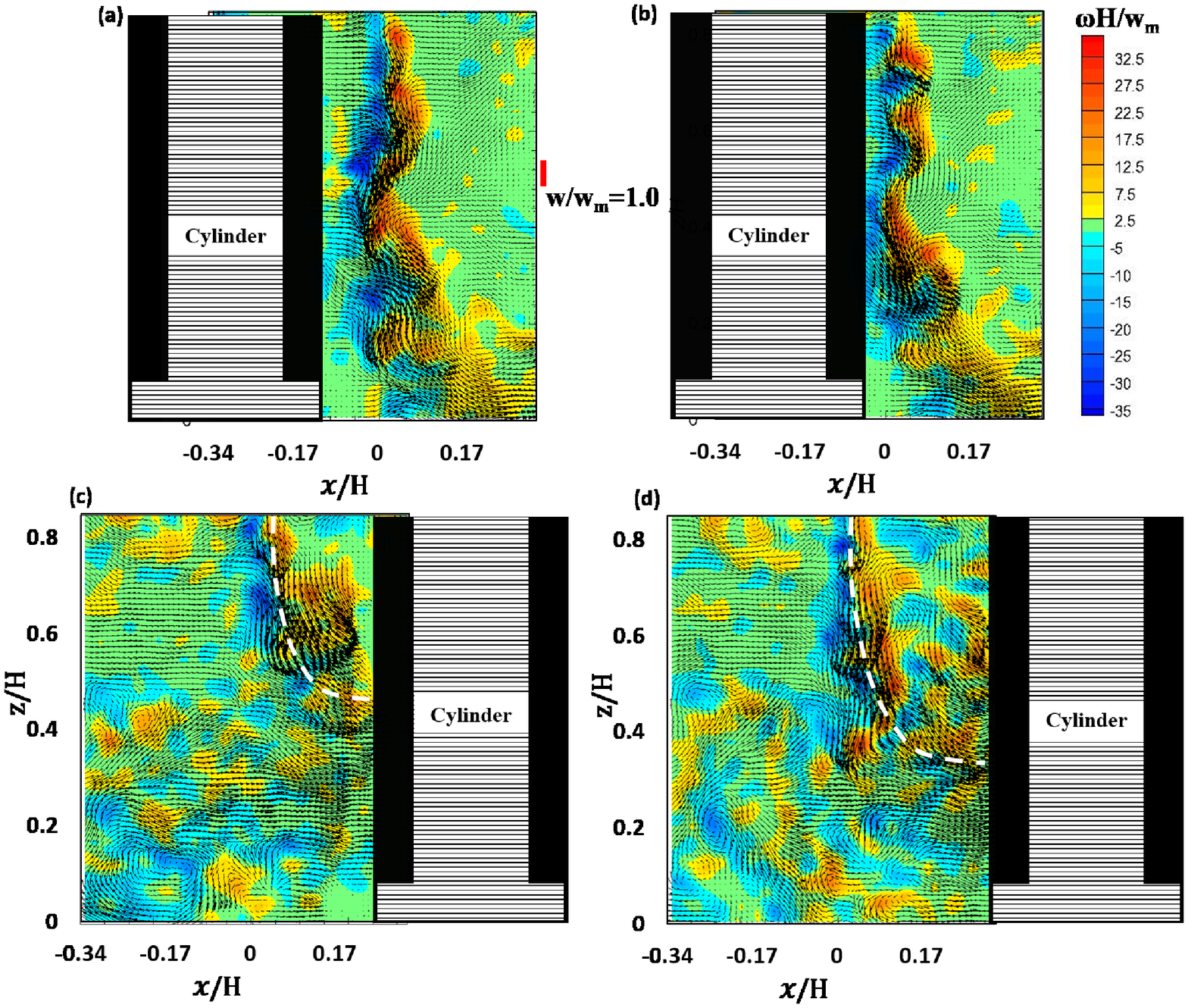}\\
\caption{ Velocity (arrows) and spanwise vorticity (colours) fields illustrating the passage of the cylinder with speed $U^*=0.42$, experiment EPIV2, through the air curtain. Figures (a) and (b) demonstrate the distortion of the curtain as the cylinder approaches, and figures (c) and (d) show the flow after the cylinder has passed through the doorway. The black space represents the area where the light is blocked by the cylinder base. The non-dimensional times are (a) $t^* =t/t_{fw}=-0.69$ and (b) $t^* = t/t_{fw}=-0.58$. (c) $t^* = t/t_{fw}=0.62$ and (d) $t^* = t/t_{fw}=0.87$. We set the time $t = 0$ as the time when the cylinder passes the jet axis. The dimensionless spanwise vorticity $\omega H/w_m$ is shown beside figure (b) and the representative length of the velocity vector $w/w_m$ is marked beside figure (a).}
\label{Fig10}
\end{figure*}

Two-dimensional PIV measurements in a vertical $x$-$z$-plane along the centreline of the channel were first conducted on the planar turbulent jet in the absence of a cylinder wake to validate the base case flow field, experiment EPIV1. Note that $z=0$ corresponds to the channel bottom and $x=0$ to the location of the ACD. For the mean velocity field, we averaged 2500 instantaneous flow fields captured at 400 f.p.s. at two vertical locations below the nozzle. As seen in figure~\ref{Fig9aa}(a), the measurements show that the planar jet was fully developed with very little variation in the mean axial velocity $\overline{w}$ at the two downstream locations. The mean axial velocity $\overline{w}$ profile in this plot is normalized by the mean centreline velocity $\overline{w}_c$ at each location and the lateral distance $x$ from the jet axis is normalized by the distance of the half velocity point $x_{\frac{1}{2}m}$. For comparison and validation, the data from \citet{Gutmark1976} and \citet{Heskestad1965} for the planar turbulent jet are also shown in the figure, with reasonable agreement between their data and ours. The axial turbulent intensity $w_{rms}$ profile, scaled by the axial mean velocity $\overline{w}_c$, is plotted against the lateral distance from the jet axis which is also non-dimensionalised by $x_{\frac{1}{2}m}$ in figure~\ref{Fig9aa}(b). The present PIV measurements are found to be in reasonable agreement with the hot-wire data of \citet{Gutmark1976} and \citet{Heskestad1965}.

Two-dimensional PIV measurements will now be presented to examine the velocity field in the vertical plane through the axis of the cylinder during the interaction of the air curtain with the cylinder wake, experiments EPIV2. As mentioned above we conducted PIV only for the pure water experiments as no quantitative infiltration flux measurements were done here for which sugar and salt are required as tracers. The flow was illuminated with a light sheet that passed through the channel base and some of the light sheet was blocked by the cylinder base. In figure \ref{Fig10}, this region is blacked out and the cylinder is marked by horizontal lines. The velocity vectors $(u,w)$ in the $x-z$-plane are represented by arrows, and the spanwise component of vorticity $\omega_y = w_x-u_z$ is represented by colours. A video of the PIV is provided in the supplementary material as \emph{`Movie1'}.

In the present case, the curtain was naturally turbulent (figure~\ref{Fig9aa}(b)) and, when the cylinder was sufficiently far away from the jet, shear layer eddies were visible on both sides of the jet. As the cylinder approached the air curtain (figure~\ref{Fig10}(a)), the streamwise velocity induced by the cylinder pushes the jet forward. This is in line with our definition of $D_{m,c}$ in equation~(\ref{eq:Dmc_def}). When the cylinder was closer to the air curtain (figure \ref{Fig10}(b)), the central core of the jet was highly perturbed and the shear layer near the cylinder was suppressed. 

The air curtain was completely disrupted when the cylinder was underneath it. After the cylinder passage, the air curtain started to re-establish but the wake velocity continued to pull it towards the cylinder (figure~\ref{Fig10}(c)).  The jet trajectory in figures~\ref{Fig10}(c) and (d) is shown by the white dotted line. The air curtain started to penetrate the cylinder wake from above and, during this time, an unhindered fluid exchange could take place below the region of the re-establishing air curtain (figure~\ref{Fig10}(c) and \ref{Fig10}(d)). 

\begin{figure*}[!ht]
\centering\includegraphics[width=1.0\linewidth]{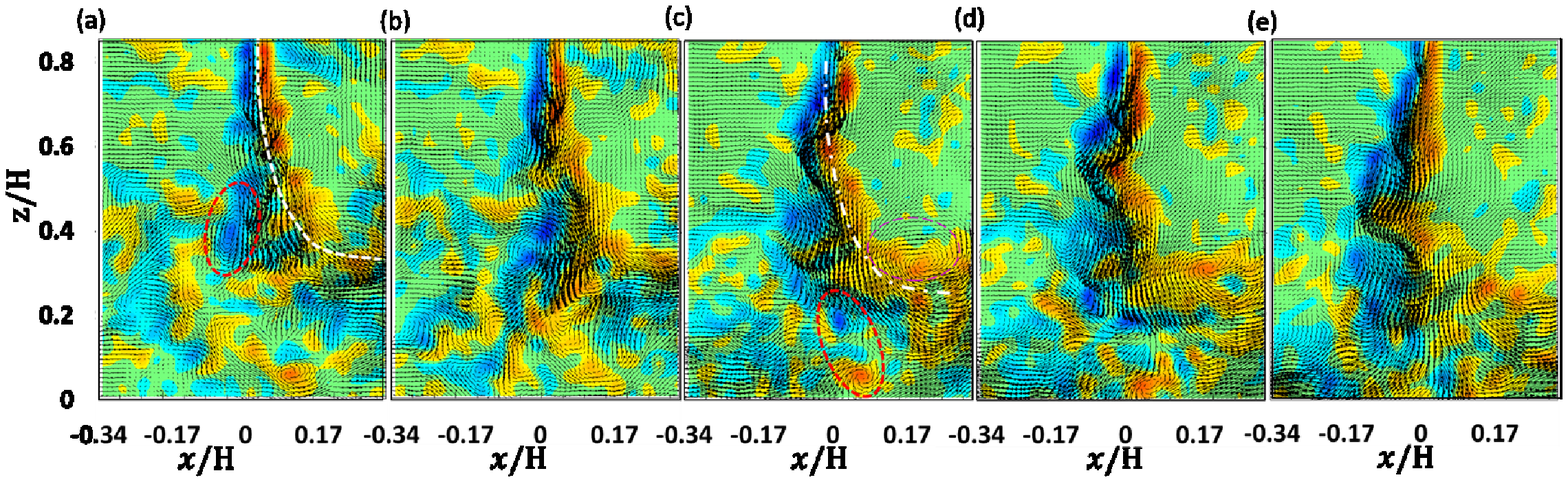}
\caption{Combined velocity and vorticity fields for a re-establishing curtain after the passage of the cylinder for the same experimental parameters as in figure~\ref{Fig10}, experiments EPIV2.  The white dashed line represents the curtain position. The evolution of the jet is downwards and deflected by the cylinder wake as seen in figure (a). At later times, the jet approaches the tank bottom and is also drawn towards the cylinder. Interaction of the re-establishing jet and the vortex pair is shown further in figures (c), (d) and (e). The vortex pair is marked by the dashed red line and a single strong vortex by the thin dashed magenta line in (c). The non-dimensional times are (a) $t^* = t/t_{fw}1.10$, (b) $t^* =t/t_{fw}=1.26$, (c) $t^* = t/t_{fw}=1.95$, (d) $t^* = t/t_{fw}=2.04$ and (e) $t^* = t/t_{fw}=2.23.$}
\label{Fig11}
\end{figure*}

In figure \ref{Fig11}, we present the meandering of the jet during the re-establishment process. The interacting vortex is marked by the dashed red line and the jet trajectory is shown by the dashed white line. The induced velocity of the marked vortex is in the clockwise direction in figure~\ref{Fig11}(a). The jet responds to the velocity induced by these vortical structures and evolves in the direction of the induced velocity. In figure~\ref{Fig11}(c), a large anti-clockwise vortical structure marked by a magenta ellipse above the establishing jet and a vortex pair marked by the dashed red ellipse can also be observed. The vortex pair induces a strong velocity at the center and the jet is then diverted towards it (figure~\ref{Fig11}(d)) and the large vortical structure draws it towards the cylinder wake. We have shown in figure~\ref{Fig11} the interaction with only two structures whereas, during the re-establishment process, the air curtain interacts in general with many such structures. The interaction of the air curtain with the vortical structures in the wake is the main cause for the meandering of the jet, which, in turn, increases the infiltration volume $V_i$ for large $U^*$ (or small $D_{m,c}$).

In section \ref{S:2}, we discussed that the conductivity measurements allow us to calculate a time-averaged flux $q_{ac}$ when the cylinder motion takes place between two fixed positions before and after the curtain. In particular, we  noted that these fixed distances were such that the flux $q_{ac}$ was elevated during a reasonable fraction of the process of the cylinder motion. The PIV measurements allow us to estimate the temporal variation of $q_{ac}(t)$ and to show that for our choice of the cylinder travel distance, the flux $q_{ac}(t)$ is indeed first elevated due to the interaction of the curtain with the cylinder and then returns back to its original value. 

\begin{figure}[!ht]
\centering\includegraphics[width=1.025\linewidth]{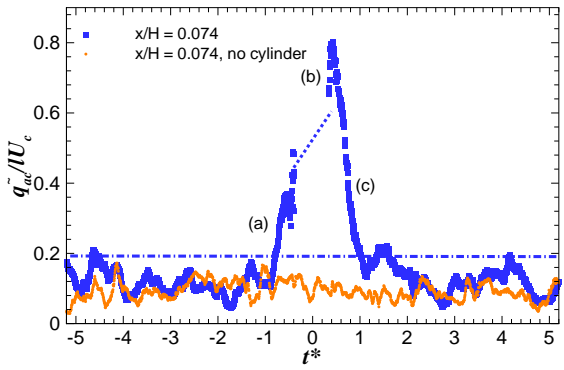}
\caption{Normalised instantaneous line flux $\tilde q_{ac}/(lU_c)$ for the position $x/H=0.08$ as a function of the normalised time $t^*=t/t_{fw}=tw_m/H$. We choose $t^*=0$ as the time when the cylinder is directly underneath the air curtain and the time axis extends for the entire process of the cylinder motion. The blue line is plotted for the case $U^*=0.42$ (experiment EPIV2) and the orange line shows the reference case of just an air curtain (experiment EPIV1) when there is no cylinder moving across it.}
\label{fig:Qac_PIV}
\end{figure}

Figure \ref{fig:Qac_PIV} shows the calculated line flux $\tilde q_{ac}/(lU_c)$ from PIV measurements for $U^*=0.42$ (experiment EPIV2) through a vertical line $x/H=0.08$ where $x=0$ corresponds to the position of the air curtain device. The time $t^*=t/t_{fw}$ is normalised by the re-establishment time of the air curtain $t_{fw}$ with $t^*=0$ chosen as the moment when the cylinder passes directly underneath the air curtain. The time axis covers the entire process of the cylinder motion, from start to finish. We note that our PIV measurements provide a visualisation of the two-dimensional flow field in the streamwise and vertical plane through the axis of the cylinder at the spanwise coordinate $y=0$. The dashed line in figure \ref{fig:Qac_PIV} corresponds to the time range when the cylinder passes through the chosen location $x/H$ and the light is blocked by the cylinder base.

As can been seen in figure~\ref{fig:Qac_PIV} the flux $\tilde q_{ac}/(lU_c)$ first increases after $t^* \sim -1$  due to the dynamic pressure of the cylinder (marked as (a) on the figure) as also seen in PIV image in figures \ref{Fig10} (a) and (b). In the time period (b) of figure~\ref{fig:Qac_PIV}, we observe that the flux increases after the passage of the cylinder caused by the transport in the cylinder wake, which is also seen in figure \ref{Fig10} (c). Finally, the flux decreases during period (c), due to the re-establishment of the air curtain as can be observed in figure \ref{Fig10}(d).

To relate the line flux $\tilde q_{ac}$ and the instantaneous flux $q_{ac}(t)$, we consider

\begin{equation}
    {\frac{q_{ac}(t)}{ldU_c}=\frac{\int_{-W/2}^{W/2} \tilde q_{ac}(y)dy}{ldU_c}.}
 \end{equation}
Now, since the interaction of the curtain with the cylinder wake is a three-dimensional process, the line flux $\tilde q_{ac}(y)$ will be generally dependent on the spanwise coordinate $y$. However, we only have the PIV measurements for the plane $y=0$. If we make a very crude approximation $\tilde q_{ac}(y)\approx \tilde q_{ac}(0)$ for $|y|<d/2$ and $\tilde q_{ac}(y)\approx 0$ otherwise, then

\begin{equation}
   {\frac{q_{ac}(t)}{ldU_c}\approx\frac{\tilde q_{ac}(y)\big|_{y=0}d}{ldU_c}=\frac{\tilde q_{ac}}{lU_c}.}
\end{equation}

The mean of the data in figure \ref{fig:Qac_PIV} is $\tilde q_{ac}/(lU_c)\sim 0.19$. The time-averaged flux $q_{ac}$ for experiment E13
calculated by means of  (\ref{eq:q_ac}) is $q_{ac}=\SI{0.35}{\liter\per\second}$ and $\frac{q_{ac}}{ldU_c}=0.23$ for EPIV2. Thus, $q_{ac}$ calculated as in (\ref{eq:q_ac}) can indeed be regarded as an approximation to the average of the temporally varying exchange flux associated with the cylinder motion. 

\section{Summary and conclusions}\label{S:4}

We have examined the contaminant transport by a cylinder moving through the air curtain which separates two iso-density sections of a channel. This models the situation in a building when a person walks along a corridor between two isothermal zones, for example a healthcare worker moving from a dirty into a clean zone in a hospital. Small-scale laboratory experiments were conducted using fresh water, salt and sugar solutions, to produce flows dynamically similar to real-scale air curtain installations. 

For a cylinder passing through an air curtain between two iso-density zones and two fixed points, we observed a reduction in infiltration flux of about 40\% as compared to the no curtain case. Furthermore, we established new definitions of the deflection modulus $D_{m,c}$ and the effectiveness $E_c$. $D_{m,c}$ is based on the momentum flux of the air curtain and the lateral momentum flux due to the cylinder and the effectiveness $E_c$ describes the ability of the air curtain to reduce the flux associated with the wake of the moving cylinder.

The cylinder disrupts the air curtain and visualisations of the jet and the cylinder wake show the infiltration of the fluid carried along within the cylinder wake underneath the disrupted jet while it is re-establishing. With an increasing cylinder speed, the entrainment in the cylinder wake also increases due to the faster wake velocity. 

We used the time-resolved two-dimensional particle image velocimetry to study the interaction of the jet and the cylinder wake. We observed that the re-establishment process of the jet is highly unsteady and the jet flaps due to the large vortical structures in the wake leading to an increased mixing across the doorway.
PIV measurements also allowed us to assess the temporal variations of the infiltration flux associated with the transient passage of the cylinder.

The effect of human passage on the contaminant transport is important in the design and the operation of hospital wards, clean rooms in chemical or pharmaceutical industries and in protecting isolation rooms for infectious and immunocompromised patients from infiltration of airborne contamination. 
Our study shows that an air curtain can help in reducing the contaminant transport in human wakes in hospital buildings for containment of patients and we suggest its usage along with negative pressure for containment wards.\\

\textbf{Acknowledgements}\\
This research has been supported by the EPSRC through grant EP/K50375/1 and Biddle BV. NKJ would gratefully acknowledge the support from PBC VATAT fellowship, Israel. We would like to thank Prof. Stuart Dalziel and Dr. Jamie Partridge for the discussion and advice during the experiments. We would also like to thank D. Page-Croft for technical support with the experimental setup.

\appendix

\section{Supplementary materials}\label{app:SuppMat}


Video caption for supplementary video 1 (Movie1.avi): Combined velocity and vorticity field from time-resolved PIV measurements of the interaction of the curtain and the cylinder for $U^*= 0.42$, experiment EPIV2. The video is played at 16 times slower than real speed. During the cylinder passage, part of the light sheet was blocked by the cylinder base, which can be seen in processed PIV images with mostly zero and few bad velocity vectors in those areas.



\bibliographystyle{jfm}
\bibliography{main.bbl}

\end{document}